\documentclass[astronomy,article,accept,pdftex,moreauthors]{mdpi}

\newcommand\Teff{\ensuremath{T_\mathrm{eff}}}
\newcommand\teff{\ensuremath{T_\mathrm{eff}}}
\newcommand\logg{\ensuremath{\log g}}
\newcommand\vsini{\ensuremath{v_{e}\sin i}}

\newcommand\met{\ensuremath{[M/H]}}
\newcommand\micro{\ensuremath{\xi_t}}
\firstpage{1} 
\pubvolume{1}
\issuenum{1}
\articlenumber{0}
\pubyear{2025}
\copyrightyear{2025}
\datereceived{} 
\daterevised{} 
\dateaccepted{August 21, 2025 } 
\datepublished{ } 
\hreflink{https://doi.org/} 

\Title{{The} 
 Use of Conditional Variational Autoencoders in Generating {Stellar Spectra} 
}

\TitleCitation{The Use of Conditional Variational Autoencoders in Generating Stellar Spectra}

\Author{{Marwan Gebran} 
 $^{1,}${*} 
\orcidA{} and Ian Bentley $^{2}$\orcidB{}}

\AuthorNames{Marwan Gebran and Ian Bentley}

\AuthorCitation{{Gebran, M.;} 
 Bentley, I.}

\address{%
$^{1}$ \quad Department of Chemistry and Physics, Saint Mary’s College, Notre Dame, IN 46556, USA\\
$^{2}$ \quad Department of Physics, Florida Polytechnic University, Lakeland, FL 33805, USA; ibentley@floridapoly.edu}

\corres{Correspondence: mgebrane@saintmarys.edu}
\abstract{We present a conditional variational autoencoder (CVAE) that generates
stellar spectra covering 4000 $\le$ {$T_{\mathrm{eff}}$} 
 $\le$ 11,000~K, $2.0 \le \log g \le 5.0$~dex, 
$-1.5 \le [\mathrm{M}/\mathrm{H}] \le +1.5$~dex, $v\sin i \le 300$ km/s, \micro\ between 0 and 4 km/s, and for any instrumental resolving powers less than 115,000. The spectra can be calculated in the wavelength range 4450--5400 \AA. Trained on a grid of \textsc{SYNSPEC} spectra, the network
synthesizes a spectrum in around two orders of magnitude faster than line‑by‑line radiative transfer. We validate the CVAE on $10^{4}$~test spectra unseen during training. Pixel‑wise statistics yield a median absolute residual of
<$1.8\times10^{-3}$ flux units with no wavelength‑dependent bias.
A residual error map across the parameters plane shows $\langle|\Delta F|\rangle<2\times10^{-3}$ everywhere, and marginal diagnostics versus $T_{\mathrm{eff}}$, $\log g$, \vsini, \micro,
and \met\ reveal no relevant trends. These results demonstrate that the CVAE can serve as a drop‑in,
physics‑aware surrogate for radiative transfer codes, enabling real‑time forward
modeling in stellar parameter inference and offering promising tools for spectra synthesis for large astrophysical data analysis.}


\begin{document}


\section{Introduction}
Since transformers were introduced~\cite{2017arXiv170603762V}, generative artificial intelligence (AI) has rapidly spread to many disciplines, including astronomy. Researchers now identify pulsar candidates with transformer‑based classifiers \citep{2024ChJPh..90..121C}. Others have created multimodal, object detection-driven augmentation models for satellite image sets \citep{2025NatSR..1512742M}. Deep‑learning frameworks can translate data from one solar instrument to another, producing homogeneous long‑term data series \citep{2025NatCo..16.3157J}. Generative AI has even been used to predict optical galaxy spectra from broadband photometry alone~\cite{2024ApJ...977..131D}. Most recently, teams have combined several generative techniques to synthesize realistic solar magnetic‑field patches and employ them as queries to locate matching structures in real observations~\cite{2024AGUFMIN01...39C}. In~stellar spectroscopy, ref.~\cite{2024MNRAS.527.1494L} developed a proof-of-concept Transformer-based model that can both predict stellar parameters from spectra and generate synthetic spectra from stellar parameters, functioning as a kind of “foundation model” for stellar spectroscopy. Their model was trained on low-resolution Gaia spectra, but~the concept demonstrates the versatility of deep learning (specifically Transformers) to serve as a forward model for spectra generation. Ref.~\cite{2021ApJ...906..130O} introduced \texttt{{Cycle-StarNet} 
}, a~hybrid generative adversarial network approach that learns to transform theoretical (synthetic) stellar spectra into spectra that look more like real observed spectra. Their goal was to bridge the “synthetic–observational gap” by correcting systematic differences between models and real data. Ref.~{\cite{2024arXiv241104750K}} 
  introduce a Transformer-based stellar foundation model, \texttt{{SpectraFM}} that treats every wavelength pixel as a token, embeds its flux plus a learned wavelength positional code, and~is pre‑trained on $\sim$90,000 synthetic APOGEE spectra to predict key stellar labels (\teff, \logg, [Fe/H], [O/Fe], [Mg/Fe]).
After a brief fine‑tune on real APOGEE data and an additional 100 star fine‑tune in a different infrared window, the~model outperforms {a neural network (NN)
} trained from scratch. Attention‑map analysis shows the Transformer naturally locks onto physically meaningful spectral lines, giving astrophysically interpretable predictions. Overall, \texttt{{SpectraFM}} demonstrates that a single, modality‑flexible foundation model can transfer knowledge across instruments and small datasets, reducing the “synthetic gap” and paving the way for few‑shot or cross‑survey spectroscopic inference in astronomy. Ref.~\cite{2022arXiv221105219C} applied physics-informed neural networks (PINNs) to solve the radiative transfer equation in a different context, specifically for supernova spectra synthesis. They used a {NN} that inherently satisfies the differential equations of radiative transfer to compute spectra, and~compared the results to a traditional radiative transfer code (\texttt{{TARDIS}}).
These are just a few recent examples to illustrate the rise in the use of AI in~astronomy.

The generation of synthetic stellar spectra is essential for many astrophysical applications including the calibration of spectroscopic surveys and the testing of stellar classification algorithms. Most stellar spectroscopy analysis techniques rely on synthetic data to be tested and constrained. Astronomers use radiative transfer codes to simulate the spectra of specific stars, planets, galaxies, and~other astronomical objects. Synthetic stellar spectroscopy relies on a limited combination of model atmospheres and radiative transfer codes. Every available radiative transfer code is usually appropriate for a specific range of stellar parameters. For~example, the~\texttt{{PHOENIX}} models~\cite{2013A&A...553A...6H} are well suited for stars having \Teff $\leq$ 12,000 K, or~\texttt{SYNSPEC} models~\cite{2011ascl.soft09022H,2017arXiv170601859H,2021arXiv210402829H}, usually used to synthesize spectra of stars with effective temperatures (\Teff) $\geq$ 4000~K. 

In the absence of direct access to the radiative transfer and model atmospheres codes, one can access specific databases that offer a selection, yet limited in sampling space, of~stellar spectra. A~comprehensive listing of the available codes and databases is described in~\cite{previous_work} and references therein. When using databases that are typically calculated with large steps in stellar parameters, the~resulting uncertainties in the derived stellar parameters become significant when compared with true~observations.

In our previous work~\cite{previous_work}, we introduced a robust methodology for constructing synthetic spectra from theoretical models. The~method was based on a combination of two distinct NNs. First, an~autoencoder is trained on a set of BAFGK synthetic data. Then, a~fully Dense NN was used to relate the stellar parameters to the Latent Space of the autoencoder. Finally, the~Fully Dense NN was linked to the decoder part of the autoencoder in {order} to build a model that uses as input any combination of the effective temperature \Teff, surface gravity \logg, projected equatorial rotational velocity \vsini, the~overall metallicity \met, and~the microturbulence velocity \micro, and~output a normalized stellar~spectrum.

Here, we extend that work of~\cite{previous_work} by exploring new deep generative models, specifically the Conditional Variational Autoencoders (CVAE), to~model and generate synthetic spectral data. This method is straightforward and requires fewer steps and preparation than the one described in~\cite{previous_work} {as it requires training only one model}. 

The paper is organized as follows: Section~\ref{DB} describes the model atmospheres and radiative transfer codes used in the database construction. Section~\ref{VAE-CVAE} details the CVAE model and its mathematical description. Section~\ref{Sec4} evaluates the generated spectra. Conclusions are outlined in Section~\ref{Sec5}.

\section{Database}
\label{DB}
 Our Training Database (DB) is built using synthetic data. The~same technique that will be presented in this paper can be applied to observational spectra that have accurate stellar parameters. It is worth noting that the idea is to reconstruct spectra similar to the ones we have in the DB, regardless of the origin or nature of these spectra. Our DB is made of a set of synthetic spectra having stellar and instrumental parameters in the range depicted in Table~\ref{stellar-parameters}. This broad range of resolving powers demonstrates that our technique is not limited to a single instrument; it can be applied to data from many different instruments and~surveys.

\begin{table}[H]
    \caption{Range of parameters used in the calculation of synthetic spectra. The~upper part of the table includes astrophysical parameters of the stars, while the lower part includes instrumental parameters. All these spectra were calculated in a wavelength range of 4450--5400 \AA. }
\newcolumntype{C}{>{\centering\arraybackslash}X}
    \begin{tabularx}{\textwidth}{CC}
    \toprule
       \textbf{Parameter}  & \textbf{Range}  \\
       \midrule
        \Teff & 4000--11,000~K \\
        \logg & 2.0--5.0~dex\\
        \vsini & 0--300 km/s \\
        \met & $-$1.5--1.5~dex \\
        \micro & 0--4 km/s \\
        \midrule

        Resolution ($\dfrac{\lambda}{\Delta \lambda}$)& 1000--115,000 \\
        \bottomrule
    \end{tabularx}
    \label{stellar-parameters}
\end{table}

\begin{figure}[H]
\begin{adjustwidth}{-\extralength}{0cm}
\centering 

    \includegraphics[scale=0.45]{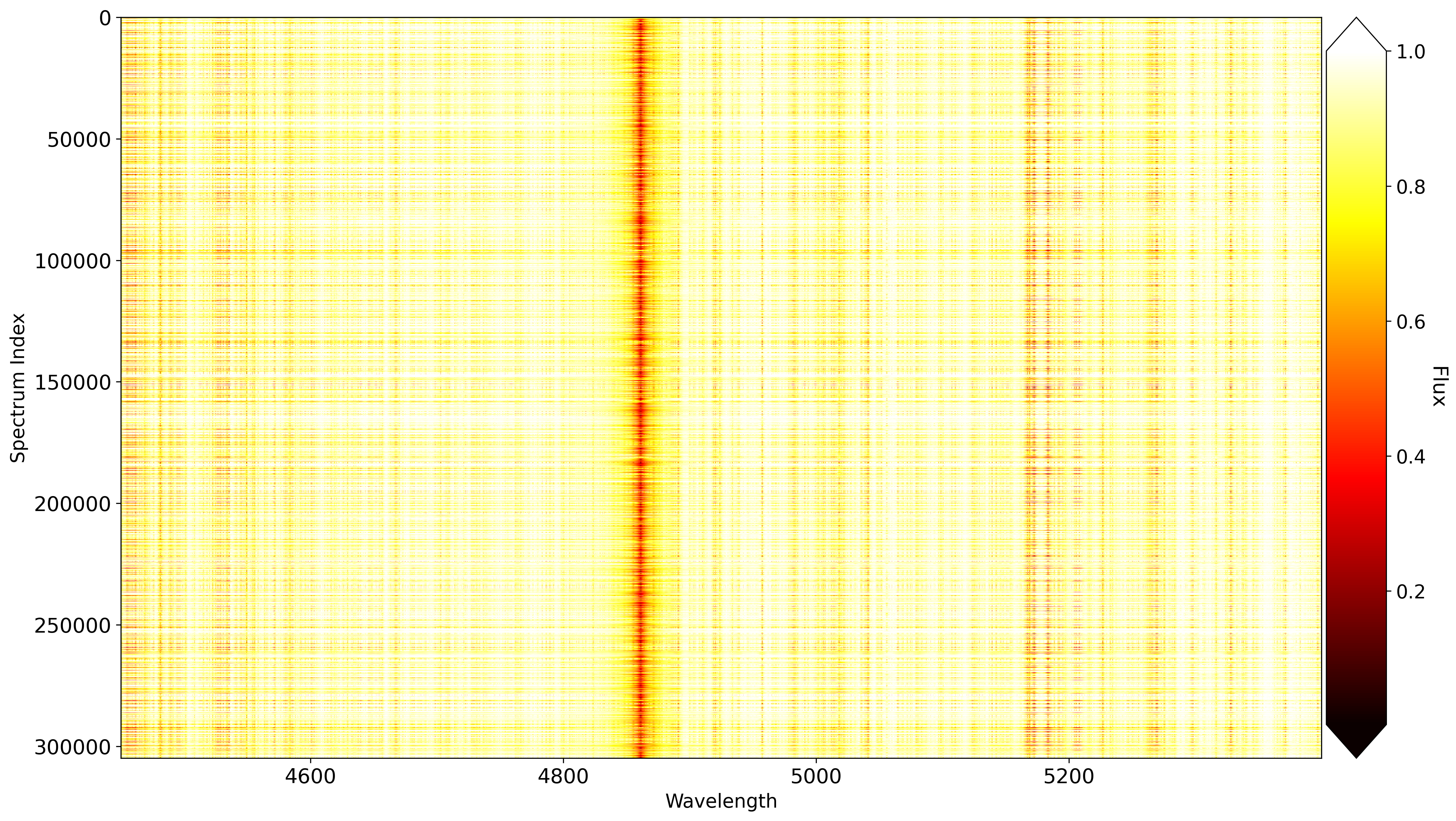}
\end{adjustwidth}
        \caption{{Color} 
 map representing the fluxes for our DB. Wavelengths are in \AA. There are 19,000~wavelength points for each~spectrum.}
    \label{colormap}
\end{figure}

Around 300,000 synthetic spectra were calculated to construct the DB. For~each spectrum, stellar and instrumental parameters were randomly selected from Table~\ref{stellar-parameters}. The~procedure for generating a synthetic spectrum is detailed in~\cite{2016A&A...589A..83G,2019OAst...28...68K,2023OAst...32..209G}. In~summary, we used \texttt{ATLAS9} (Kurucz,~\cite{1992IAUS..149..225K}) to calculate line-blanketed model atmospheres for this work. These are LTE plane parallel models that assume hydrostatic and radiative equilibrium. We used the Opacity Distribution Function (ODF) of~\cite{2003IAUS..210P.A20C}. For~stars cooler than 8500 K, we incorporated convection using Smalley’s prescriptions~\cite{2004IAUS..224..131S} and the mixing length theory. The~mixing length parameter was 0.5 for 7000 K $\leq$ \Teff~$\leq$~8500 K and 1.25 for \Teff~$\leq$ 7000 K. The~synthetic spectra grid was computed using \texttt{SYNSPEC} \cite{2011ascl.soft09022H} according to the parameters described in Table~\ref{stellar-parameters}. We scaled the metallicity, with~respect to the Grevesse \& Sauval solar value~\cite{1998SSRv...85..161G}, from~$-$1.5~dex up to +1.5 dex. The~metallicity is calculated as the abundance of elements heavier than Helium. The~change in metallicity consists of a change in the abundance of all metals with the same scaling factor. The~synthetic spectra were computed from 4450 \AA\ up to 5400 \AA\ with a wavelength step of 0.05 \AA. This range contains many moderate and weak metallic lines in different ionization stages. These weak metallic lines are sensitive to \vsini, \met, and~\micro, while the Balmer line is sensitive to \teff\ and \logg. 
The linelist used in the synthetic spectra calculation was constructed from Kurucz database and modified with updated atomic data as explained in~\cite{2023OAst...32..209G}. Figure~\ref{colormap} shows a colormap of the full database. The~fluxes of the normalized spectra are shown as the intensity in the colorbar. The~strong absorption line around 4861 \AA\ corresponds to the Balmer $H_\beta$ line. This range of wavelength contains many lines with different information on the chemical abundance of many metals such as Mg, Si, Ca, Sc, Ti, Cr, Mn, Fe, Ni, and~Zr, among~others. These chemical elements have different ionization stages and are sensitive to the stellar and instrumental~parameters.

\unskip

\section{Variational~Autoencoder}
\label{VAE-CVAE}
Variational Autoencoders (VAEs) are a class of deep generative models that learn to encode data into a lower-dimensional latent space and then decode it back to reconstruct the original data. Unlike traditional autoencoders, VAEs impose a probabilistic structure on the latent space by encoding inputs to probability distributions rather than to fixed points~\cite{VAE}. This probabilistic approach allows VAEs to serve as generative models capable of producing {new}, realistic data samples. A~standard VAE consists of two primary components, an~encoder network that maps input data $x$ to a distribution in latent space, typically parameterized by means $\mu$ and log-variances $\log\sigma^2$, and~adecoder network that maps samples from the latent space back to the data space, reconstructing the original~input.

The VAE is trained to minimize two objectives simultaneously: The reconstruction loss that ensures decoded outputs closely match the inputs and the Kullback--Leibler (KL) divergence that ensures the encoded latent distributions approximate a prior distribution, typically a standard normal distribution $\mathcal{N}(0, I)$.

\subsection{Conditional Variational~Autoencoders}
Conditional Variational Autoencoders (CVAEs) extend the VAE framework by incorporating conditioning information. In~a CVAE, both the encoding and decoding processes are conditioned on additional variables $c$. This conditioning enables the model to generate data with specific desired properties or characteristics. Typically, the~conditioning variable $c$ contains the values of the stellar and instrumental~parameters. 

Formally, while a VAE models the marginal likelihood $p(x)$, a~CVAE models the conditional likelihood $p(x|c)$. The~encoder in a CVAE learns to approximate the posterior $q_\phi(z|x,c)$ rather than $q_\phi(z|x)$, and~the decoder learns the conditional generative model $p_\theta(x|z,c)$.

\clearpage
In our specific application, we employ a CVAE to model and generate synthetic stellar spectra conditioned on physical stellar parameters. This approach offers several advantages. It provides a data-driven model for spectrum synthesis that complements traditional physics-based models, it enables rapid generation of spectra once trained, facilitating large-scale analyses, and~it allows for exploration of the continuous space of stellar parameters through the conditioned generative~process.

Our CVAE implementation uses the 6 parameters of Table~\ref{stellar-parameters} as conditioning variables. Its architecture consists of two main blocks, An encoder and a decoder network. The~encoder takes two inputs, the~stellar spectrum represented as a 1D array with shape $(n_{\text{wav}}, 1)$, where $n_{\text{wav}}$ is the number of wavelength points which is 19,000 in our case, and~the conditioning stellar parameters vector with shape {$(6,)$} 
 containing $T_{\text{eff}}$, $\log g$, \vsini, [M/H], \micro, and~the resolving~power.

The encoder processes these inputs through several dense layers and outputs the mean vector $\mu$ of the latent distribution and the log-variance vector $\log\sigma^2$ of the latent distribution. This will allow us to sample a latent vector $z$ from this distribution in the \mbox{form of:}
\begin{equation}
z = \mu + \exp(0.5 \cdot \log\sigma^2) \odot \epsilon
\end{equation}
where $\epsilon$$\sim$$\mathcal{N}(0, I)$ is a random noise~vector.

Next, the~decoder takes the sampled latent vector $z$ from the encoder and the same conditioning stellar parameters vector. Through several dense layers, the~decoder reconstructs the original spectrum with shape $(n_{\text{wav}}, 1)$.

In a similar way to traditional VAE, the~model is trained by minimizing the loss function:
\begin{equation}
\mathcal{L} = \mathcal{L}_{\text{reconstruction}} + \mathcal{L}_{\text{KL}}
\end{equation}
where
\begin{itemize}
\item $\mathcal{L}_{\text{reconstruction}} = \text{MSE}(x, \hat{x})$ is the mean squared error between the original spectrum $x$ and the reconstructed spectrum $\hat{x}$.
\item $\mathcal{L}_{\text{KL}} = -\frac{1}{2} \sum_{j=1}^J (1 + \log\sigma^2_j - \mu^2_j - \exp(\log\sigma^2_j))$ is the KL divergence term that regularizes the latent space.
\end{itemize}

Once trained, our CVAE can generate synthetic stellar spectra for any given set of physical parameters through the following~process:
\begin{itemize}
\item Normalize the desired stellar parameters using the previously calculated normalization factors:
\begin{equation}
c_{\text{normalized}} = \frac{c - c_{\text{min}}}{c_{\text{max}} - c_{\text{min}}}
\end{equation}
\item Sample a random vector $z$ from the standard normal distribution $\mathcal{N}(0, I)$.

\item Feed $z$ and $c_{\text{normalized}}$ to the decoder to generate a normalized synthetic spectrum:
\begin{equation}
\hat{x}_{\text{normalized}} = \text{decoder}(z, c_{\text{normalized}})
\label{generation_equ}
\end{equation}

\item Denormalize the spectrum to obtain physical flux values (the generated spectrum):
\begin{equation}
\hat{x} = \hat{x}_{\text{normalized}} \cdot (x_{\text{max}} - x_{\text{min}}) + x_{\text{min}}
\end{equation}
\end{itemize}

\subsection{Model~Architecture}

In our implementation, we used the following architecture~specifications:
\begin{itemize}
\item An input layer containing the spectrum of dimension 19,000 combined with a conditional vector containing the stellar and instrumental parameters of dimension 6.
\item Latent space of dimension 100. 
\item Encoder network: Dense layers with 4000, 2000, and~1000 units with ReLU activations.
\item Decoder network: Dense layers with 1000, 2000, and~4000 units with ReLU activations, followed by a final layer with sigmoid activation.
\item Training parameters: Adam optimizer with a dynamical learning rate, a~batch size of 512, and~early stopping based on reconstruction loss with patience of 50.
\end{itemize}

\begin{figure}[H]

\begin{adjustwidth}{-\extralength}{0cm}
\centering 

    \includegraphics[width=0.8\linewidth]{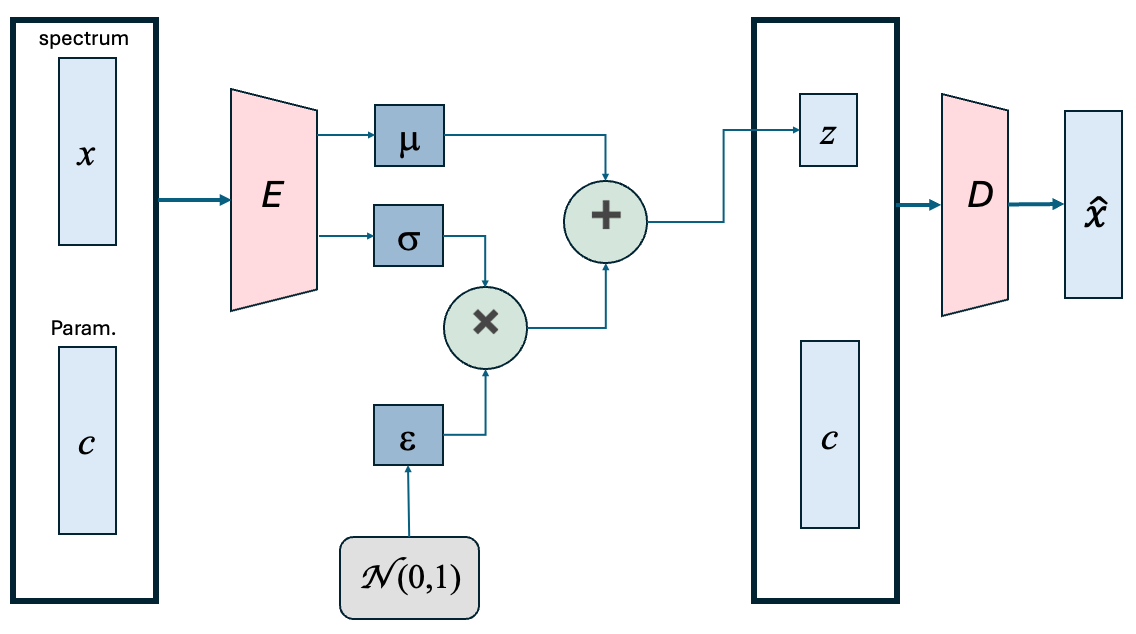}
\end{adjustwidth}
    \caption{Flowchart of the CVAE used in this work. Boxes labelled $E$ and $D$ are the encoder and decoder; $\mu$ and $\sigma$ are the latent‑space mean and variance; $z$ is the 100‑dimensional latent vector; $c$ is the 6‑element conditioning vector (\teff, \logg, \vsini, \met, \micro, Resolution). $x$ is the input spectrum and $\hat{x}$ is the output spectrum. The~flowchart is inspired by a similar chart in~\cite{2021arXiv210410093V}.}
    \label{architecture}
\end{figure}

The choice of this network is based on the size of the data and on several trials that lead to an optimization of the time and errors. The~training for the 300,000 spectra on a 52--2.1 Ghz core computer with 258 GB RAM and 48 GB Nvidia RTX A6000 graphic car took around 420 h. A~flowchart of the network is presented in Figure~\ref{architecture} in which the encoder $E$ contains the succession of the three dense layers of 4000, 2000, and~1000 nodes, respectively. In~the same way, the~decoder $D$ contains a successions of 3 dense layers of 1000, 2000, and~4000 nodes, respectively.

\subsection{Spectra~Generation}
\begin{figure}[H]
\begin{adjustwidth}{-\extralength}{0cm}
\centering 

    \includegraphics[width=0.9\linewidth]{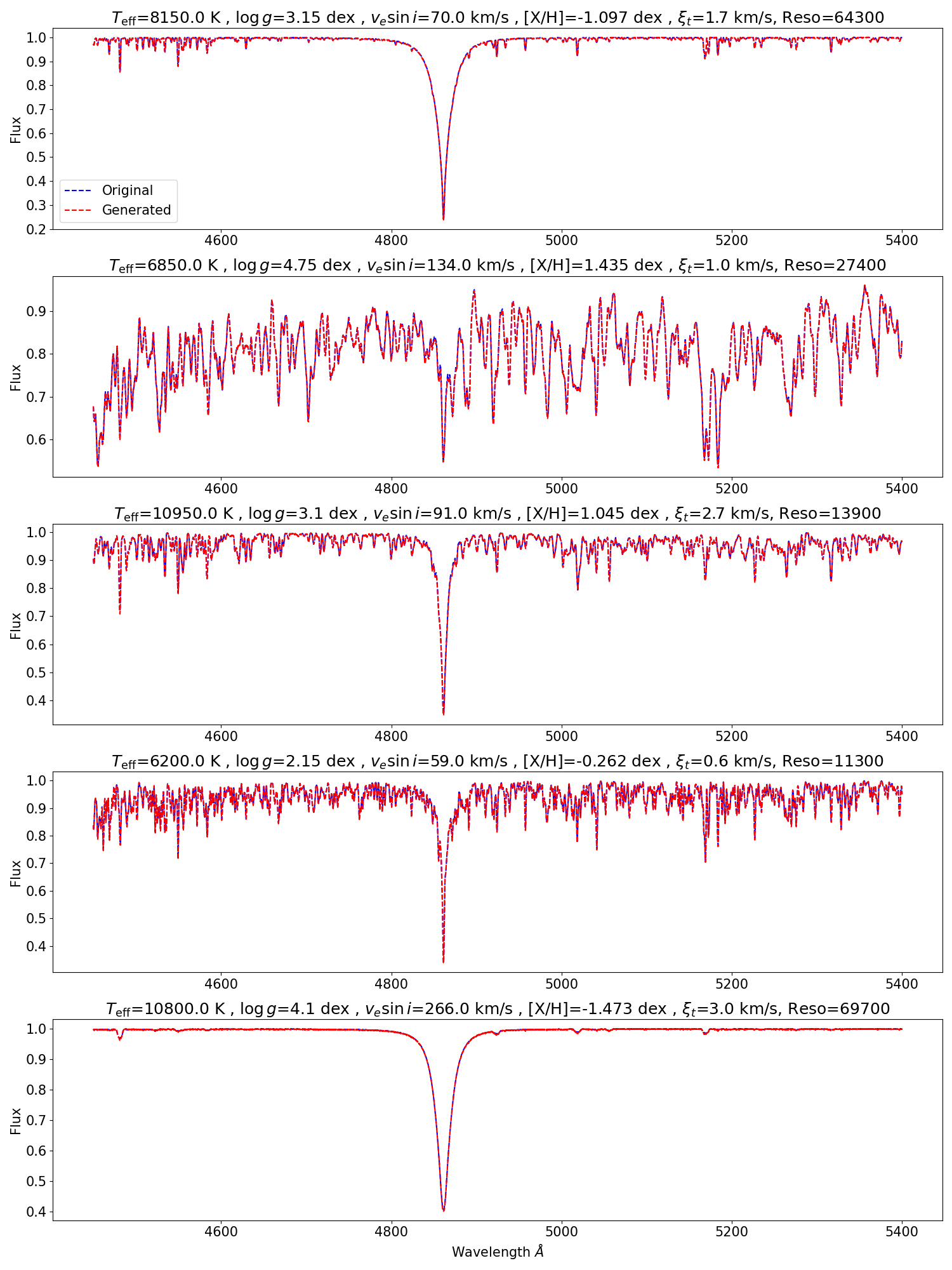}
\end{adjustwidth}
    \caption{{Generated} 
 spectra (red) compared to the ones calculated using \texttt{{SYNSPEC}} ({blue}) having the same combination of stellar parameters and resolution. Each spectrum has a different combination of stellar parameters \Teff, \logg, \vsini, \met, \micro, and~resolution.}
    \label{genvssynspec}
\end{figure}

Once the model is trained, the~generation of spectra are done using the decoder part and by choosing randomly a set of stellar and instrumental parameters that play the role of the conditional variable $c$. This is done using Equation~(\ref{generation_equ}), defined previously. The~generated spectra are computed using any configuration of parameters {within the bounds} of Table~\ref{stellar-parameters} and not necessarily a combination that exists in the DB. Figure~\ref{genvssynspec} represents a sample of randomly generated synthetic spectra (in red dashed line) displayed with the ones calculated using the radiative code \texttt{{SYNSPEC}} for the same parameters (in blue). This visual inspection shows that the CVAE is capable of reproducing all the features of the original spectra in this wavelength region. A~more quantitative assessment of the generated spectra quality is performed in the next section. Furthermore, for that, we have constructed a generated database of 15,000 generated spectra that we will be~analyzing. 

\unskip

\section{Determination of~Parameters}\label{Sec4}
A quantitative assessment of the quality of the generated spectra is carried out using a technique that has been previously tested on the original synthetic spectra. By~applying the same technique to both the synthetic spectra and the generated spectra, we can check if we observe identical results. If~this is the case, this confirms that the generated spectra cannot be distinguished from the original~ones. 

We have performed a parameter determination and checked the inferred accuracies between the synthetic ones and the generated ones. The~determination of the stellar and instrumental parameters are done according to the work of~\cite{2022OAst...31...38G,2023OAst...32..209G}. It consists of developing a NN that is trained on 6 parameters (\teff, \logg, \vsini, \met, \micro, and~resolution). The~network involves a preprocessing of the original spectra using a {Principal Component Analysis (PCA)} transformation. The~input data consists of a matrix of synthetic data having a dimension of 600,000 spectra $\times$ 19,000 wavelength points. This matrix is reduced to 600,000 spectra $\times$ 25 PCA coefficients. As~explained in~\cite{2023OAst...32..209G}, this step is optional but recommended to increase the speed of the calculations. {The} choice of the number of coefficients is regulated by the reconstructed error from PCA (see~\cite{2015A&A...573A..67P,2023OAst...32..209G} for more details). The 600,000 spectra are in fact the original ones of the DB augmented with the noisy ones according to~\cite{2022OAst...31...38G}. We followed the same technique as in~\cite{2022OAst...31...38G,2023OAst...32..209G} by using the same DB and apply a random Gaussian noise, with~a signal to noise ratio, SNR, between~5 and 300, to~each spectrum of the DB. 
The choice of the number of coefficients is regulated by the PCA reconstructed error (see~\cite{2015A&A...573A..67P,2023OAst...32..209G} for more details).

\subsection{Accuracy of the Stellar~Parameters}

We have performed a parameter determination and compared the inferred accuracies between synthetic and generated spectra. The~determination of stellar and instrumental parameters is based on the work of~\cite{2022OAst...31...38G,2023OAst...32..209G}. We developed a NN trained on six parameters: \teff, \logg, \vsini, \met, \micro, and~resolution.  Using the same architecture of~\cite{previous_work}, we have constructed a model that related the 25 PCA coefficients to the 6 stellar and instrumental parameters as displayed in Table~\ref{archite-stellar-param}.

\begin{table}[H]
\caption{{Architecture} 
 of the Fully Connected Neural Network used to relate the 25 PCA coefficients to the 6 stellar and instrumental~parameters.}
\label{archite-stellar-param}
\newcolumntype{C}{>{\centering\arraybackslash}X}
    \begin{tabularx}{\textwidth}{LCC}
   \toprule
 \textbf{Layer} & \textbf{Characteristics} & \textbf{Activation Function} \\
\midrule
 {Input} & PCA coefficient (25 data points per spectrum) & - \\
 Hidden & 5000 neurons & ReLU \\
 Hidden & 2000 neurons & ReLU \\
 Hidden & 1000 neurons & ReLU \\
Hidden & 64 neurons & ReLU \\
{Output} & Stellar Parameters (6 data points per spectrum) & - \\
\bottomrule
\end{tabularx}
\end{table}

The data (i.e., the 600,000 spectra) has been divided in {70-15-15\%} 
 for training, validation, and~testing, respectively. The~optimizer used is ``Adamax'' combined with a Mean Squared Error (MSE) loss function. We have also used a batch size of 1024. Once the model is trained and the loss function minimized, the~results are displayed in Table~\ref{results} in which we have calculated the root of the MSE between the original and the derived stellar and instrumental parameters for the training, validation, test, and~generated data. The~purpose of this test is to show that the generated data represents the same features as the ones calculated using the radiative transfer code. Therefore, if~the derived accuracies for the generated data are in the same order as the ones for the test data, it shows that our approach is capable of reproducing spectra as accurate as the radiative transfer code. We are not assessing the technique for deriving the parameters, the~parameters derivations, this has been done in~\cite{2022OAst...31...38G,2023OAst...32..209G,previous_work,2025OAst...3440010G}.

\begin{table}[H]
\caption{Derived accuracies of the stellar parameters for the training, validation, test, and~generated database. These values are the $\sqrt{\text{MSE}}$ between the original and the derived stellar and instrumental~parameters.}
\label{results}
\newcolumntype{C}{>{\centering\arraybackslash}X}
    \begin{tabularx}{\textwidth}{CCCCC}
   \toprule
 \textbf{Parameter} & \textbf{Training} & \textbf{Validation} & \textbf{Test} & \textbf{Generated} \\
\midrule
\teff\ (K)&30 &45 &60 &65 \\
\logg\ (dex) & 0.04& 0.04&0.04 &0.04 \\
\vsini\ (km/s) &3.0 &5.1 &6.2 &6.1 \\
\met\ (dex) &0.030 &0.035 & 0.029&0.030 \\
\micro\ (km/s) &0.08 &0.10 &0.08 &0.09 \\
\bottomrule
\end{tabularx}
\end{table}

\begin{figure}[H]
\begin{adjustwidth}{-\extralength}{0cm}
\centering 

    \includegraphics[scale=0.37]{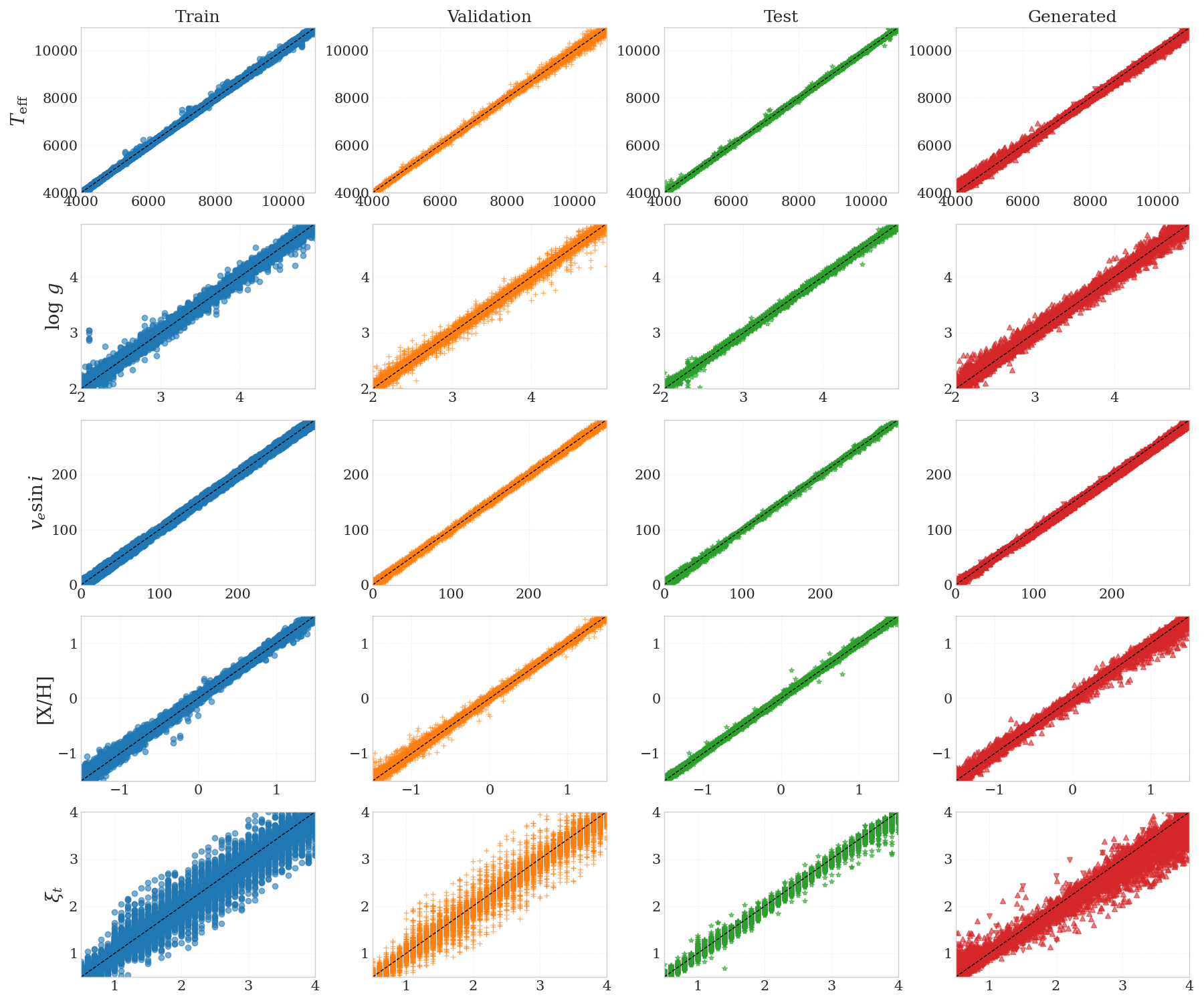}
\end{adjustwidth}
    \caption{{Predicted} 
 stellar parameters as a function of the actual ones for the training, validation, test, and~generated datasets for \teff, \logg, \vsini, \met, and~Resolution. The~scatter around the $y=x$ line is represented quantitatively in Table~\ref{results}. }
    \label{evaluation-plot}
\end{figure}

Figure~\ref{evaluation-plot} displays the predicted stellar parameters as a function of the original ones for the training, evaluation, test, and~generated dataset. Table~\ref{results} and Figure~\ref{evaluation-plot} show the similarity in the behavior of the results for the test and generated~databases.

\unskip

\subsection{Heat‑Map of Residuals over Parameter~Space}\label{sec:heatmap}

To visualize how the CVAE behaves on the stellar parameter grid, we calculate 2 new grids of spectra. These 2 grids have the same parameters except that one is calculated using \texttt{{SYNSPEC}} and the other one using the CVAE. It is worth mentioning that the gain in computation time between the CVAE and \texttt{{SYNSPEC}} is around $\sim$50\textbf{$\times$}. We then have calculated, for~each spectrum, the~wavelength‑averaged absolute residual $\langle|\Delta F|\rangle$ between the two grids in the following way
\begin{equation}
\langle|\Delta F|\rangle = \frac{1}{N_{\lambda}}\sum_{i=1}^{N_{\lambda}} \left|F_{\mathrm{gen}}(\lambda_i) - F_{\mathrm{syn}}(\lambda_i)\right|
\end{equation}
where $F_{\mathrm{gen}}(\lambda_i)$ represents the generated spectrum at the wavelength $\lambda_i$ and $F_{\mathrm{syn}}(\lambda_i)$ represents the \texttt{{SYNSPEC}} spectrum at the same wavelength $\lambda_i$. The~average is performed over the $N_{\lambda}=19,000$ wavelength points. A~smaller $\langle|\Delta F|\rangle$ indicates a closer match to the radiative transfer~solution.

\begin{figure}[H]

\includegraphics[width=0.97\textwidth]{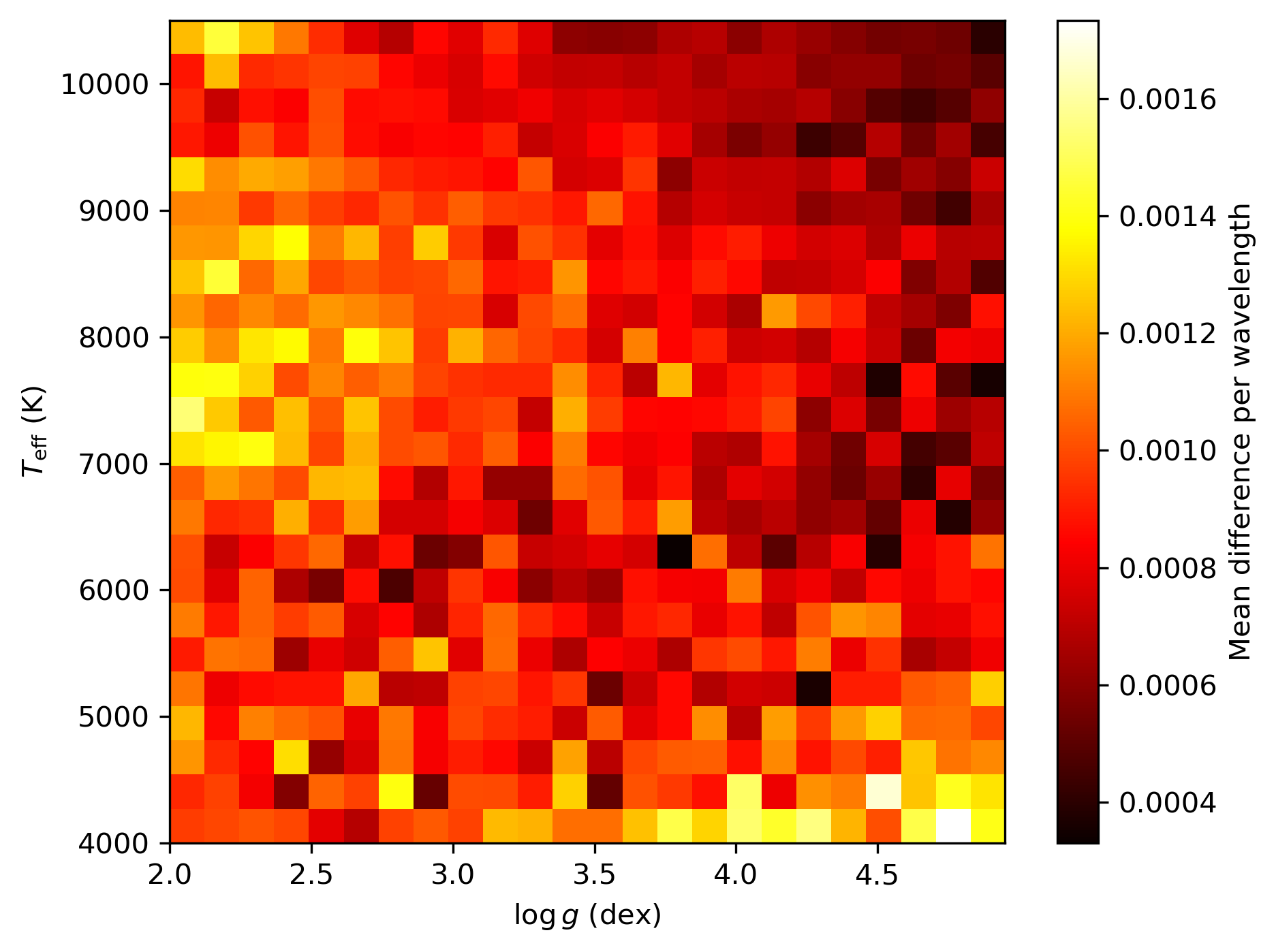}
\caption{{Mean} 
 absolute residual $r$ between CVAE‑generated and \texttt{SYNSPEC} spectra across the $T_{\mathrm{eff}}$--$\log g$ grid.  White squares indicate bins with fewer than five test spectra.  Overall, }
\label{fig:heatmap}
\end{figure}

Each spectrum being associated with its effective temperature $T_{\mathrm{eff}}$ and surface gravity $\log g$, we have generated a heat map that represents the residual in a $T_{\mathrm{eff}}$--$\log g$ plane. In~each cell we average the residuals of all spectra falling into that bin.
 The main purpose of this task is to show that there is no specific trends in the residual and therefore no biases in the CVAE spectra generation procedure. The~heat map is presented in Figure~\ref{fig:heatmap} in which we show that, in~overall, the~network achieves sub‑percent accuracy throughout the region spanned by our data. The~CVAE reproduces \texttt{SYNSPEC} with \mbox{$\langle|\Delta F|\rangle \lesssim 1.8\times10^{-3}$} across the bulk of the grid.  Slightly higher residuals appear only at cool high gravity corner where spectral lines are intrinsically denser. No trend can be found in this data, and~this is true in all the combinations of 2D planes confirming that  the latent manifold learned by the CVAE interpolates smoothly in all stellar~labels.

\unskip

\subsection{Marginal Residuals Versus Stellar~Parameters}
\label{sec:marginal_residuals}

\begin{figure}[H]
\vspace{-6pt}
    \includegraphics[width=0.99\textwidth]{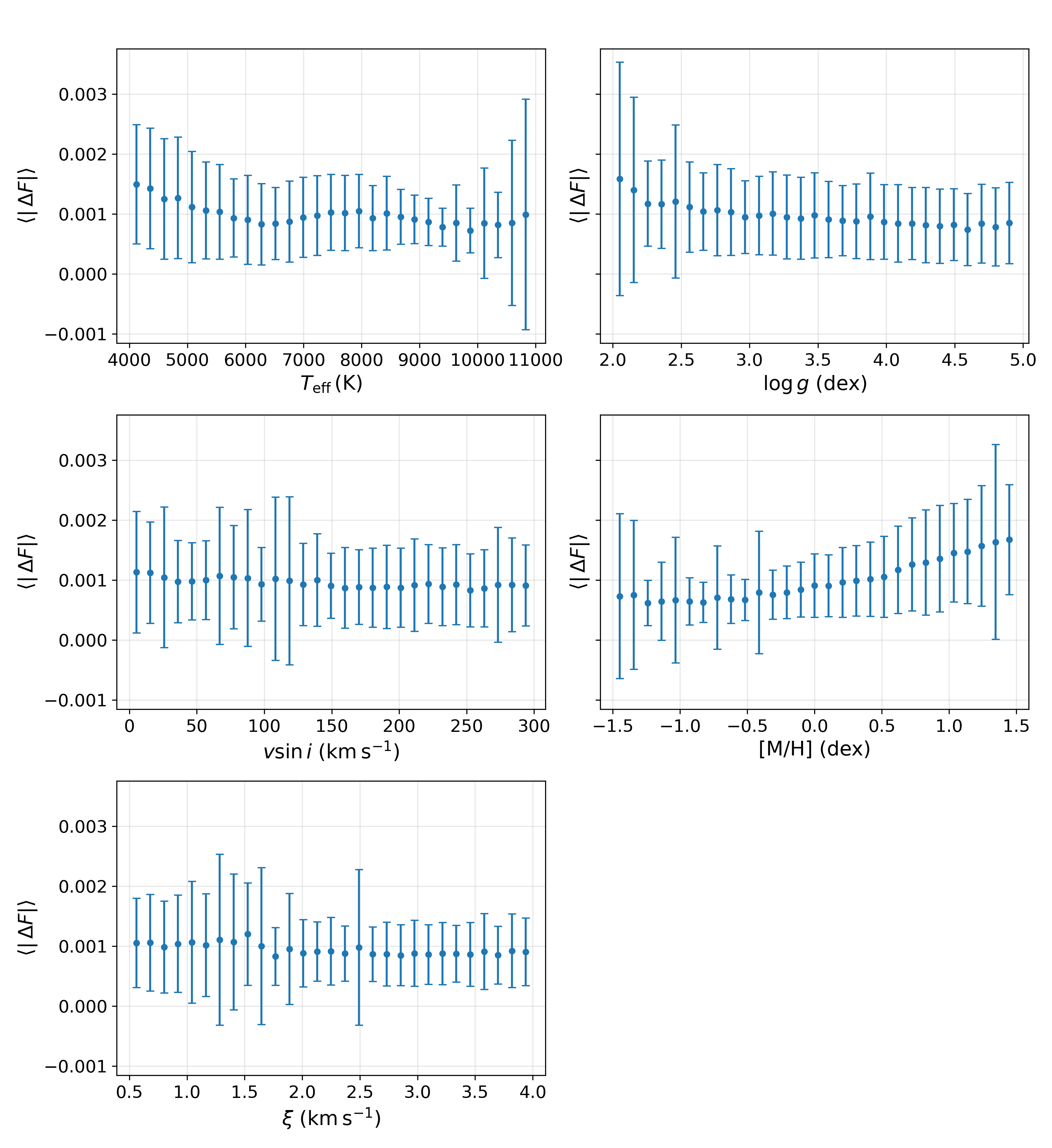}
    \caption{{Binned} 
 mean absolute residual
             $\langle|\Delta F|\rangle$ (points) and
             $1\sigma$ scatter (error bars) as a function of
             $T_{\mathrm{eff}}$, $\log g$, \vsini, $[\mathrm{M}/\mathrm{H}]$, and~\micro.
             No significant trend is observed in any parameter,
             corroborating the two–dimensional analysis of
             Section~\ref{sec:heatmap}.}
    \label{fig:marginal_residuals}
\end{figure}

To complement the two–dimensional heat‑map of Section~\ref{sec:heatmap}, Figure~\ref{fig:marginal_residuals} displays the one dimensional behavior of the wavelength averaged absolute residual, $\langle|\Delta F|\rangle$,  as~a function of four key stellar labels. The~data are binned into equal‑width intervals; solid points mark the mean residual in each bin and the error bars denote the $1\sigma$ scatter of the spectra falling in that~interval.

Although correlation tests return highly significant p‑values owing to the large sample size, the~effect sizes remain small: \teff, \logg, \vsini, and~\micro\ explain $< 4$\% of the variance in $\langle|\Delta F|\rangle$, and~metallicity explains $\sim$10\%. Even at the extremes of parameter space the mean residual never exceeds $2\times 10^{-3}$, well below the pixel noise of spectra for a SNR of around a 100. We therefore regard the residual trends as astrophysically negligible; the CVAE reproduces \texttt{SYNSPEC} to better than 0.2\% across the full~grid.

This analysis combined with the previous scatters and acuracies on the stellar parameters, concludes
that the CVAE introduces no systematic bias with respect to any of the stellar labels. The~network therefore serves as a reliable, ``physics‑aware'' surrogate for \texttt{{SYNSPEC}} over the full stellar parameter range explored in this~work.

\unskip


\section{Conclusions and Future~Work}\label{Sec5}

We have demonstrated that a CVAE trained on a grid of \texttt{{SYNSPEC}} spectra can reproduce the underlying radiative‑transfer physics to sub‑percent accuracy while delivering spectra in a fast and reliable way. The~derived parameters of the generated spectra reveal a similar accuracy to those of spectra synthesized with \texttt{{SYNSPEC}}. 

Across $10^{4}$ independent test cases the median wavelength‑normalised
residual is $\sim$$10^{-3}$ and never exceeds $1.8\times10^{-3}$ over the explored parameter space of Table~\ref{stellar-parameters}. Residual heat‑maps and one‑dimensional marginal diagnostics reveal no astrophysically significant dependence on any individual~label.

Because the CVAE learns the mapping between stellar labels and flux
rather than the internal radiative‑transfer equations, the~same architecture
can be trained on any large collection of synthetic spectra, regardless of which tools and codes are used, including the following:

\begin{itemize}
  \item Radiative transfer codes: \texttt{{SYNSPEC}}, \texttt{{TURBOSCPECTRUM}}~\cite{2008A&A...486..951G}, \texttt{{MOOG}}~\cite{2012ascl.soft02009S}$\cdots$
  \item Model atmospheres: \texttt{{ATLAS}}, \texttt{{TLUSTY}}~\cite{2011ascl.soft09022H,2017arXiv170601859H,2021arXiv210402829H}, \texttt{{PHOENIX}}~\cite{2013A&A...553A...6H}$\cdots$ 
  \item Spectral window: ultraviolet, optical, infrared, or~a combination thereof
  \item Resolving power: high‑resolution echelle down to broad‑band photometric passbands.
\end{itemize}

Our next priority is to diversify the training corpus.  This can be done by merging grids from multiple radiative transfer sources covering a broader range of physical parameter used in stellar population studies. Within~that expanded database we plan to introduce individual modified chemical abundances. In~parallel, we will retrain the network over an extended wavelength domain and not only limited to the one tested in this current work. To~further suppress the mild metallicity trend detected in this work we will experiment with physics‑informed losses that penalize deviations in equivalent widths, Balmer line wings, and/or continuum~level. 

Moreover, because~training only requires input–output pairs, the~method
is equally applicable to real high-SNR observations. One can fine‑tune the network on homogenized, normalized spectra in the same wavelength range. Once trained, the~CVAE becomes a standalone spectra generator that eliminates the hassle of using model atmospheres and radiative transfer codes (combined with their compilers) or the use of online databases that are limited in resolution and parameter space. An~example of real observations database in a large wavelength range is \texttt{{Melchiors}} \cite{2024A&A...681A.107R}. This database combines around 3250 high-SNR spectra of O to M stars between 3900 and 9000 \AA\ with a spectral resolution of 85,000.

By decoupling synthetic spectrum generation from first principles radiative transfer, our CVAE framework transforms spectrum synthesis from a computational bottleneck into a millisecond‑scale, differentiable operation. A~differentiable surrogate lets users back‑propagate through spectral synthesis, enabling gradient‑based Bayesian inference (e.g., MCMC) and automatic uncertainty propagation, neither of which is possible with traditional radiative‑transfer codes. We anticipate that such learned surrogates will become a standard infrastructure component in stellar astrophysics, enabling the astronomical community to generate high‑fidelity spectra on demand, without~invoking traditional radiative‑transfer~codes.



\vspace{6pt} 




\authorcontributions{{~} 
``Conceptualization, M.G. and I.B.; methodology, M.G.; software, M.G.; validation, I.B.; formal analysis, M.G.; investigation, M.G.; resources, M.G and I.B.; data curation, M.G.; writing---original draft preparation, M.G.; writing---review and editing, M.G. and I.B; visualization, M.G.; supervision, M.G.; project administration, M.G.; All authors have read and agreed to the published version of the manuscript.'', please turn to the  \href{http://img.mdpi.org/data/contributor-role-instruction.pdf}{CRediT taxonomy} for the term explanation. Authorship must be limited to those who have contributed substantially to the work~reported.
}


\funding{This research received no external~funding.}

\acknowledgments{{M.G.} 
 acknowledges Saint Mary's College for providing the necessary computational power for the success of this project. }


\conflictsofinterest{The authors declare no conflict of~interest.} 





\begin{adjustwidth}{-\extralength}{0cm}
\reftitle{References}

\PublishersNote{}
\end{adjustwidth}
\end{document}